\documentclass[sigconf,nonacm]{acmart}

\usepackage{booktabs}
\usepackage{multirow}
\usepackage{amsmath}
\usepackage{xcolor}
\usepackage{subcaption}
\usepackage{xspace}
\usepackage{graphicx}
\usepackage[utf8]{inputenc}
\usepackage{enumitem}
\usepackage{algorithm}
\usepackage{algpseudocode}

\algrenewcommand\algorithmicrequire{\textbf{Input:}}
\algrenewcommand\algorithmicensure{\textbf{Output:}}
\algrenewcommand\textproc{\textnormal}

\algnewcommand\algorithmicvariables{\textbf{Variables:}}
\algnewcommand\Variables{\item[\algorithmicvariables]}

\title{GoalEvolve: From Handcrafted Algorithm Priors to Goal-Driven Evolution of Physical Design Algorithms}

\author{Haixu Liu}
\affiliation{%
  \institution{Fudan University}
  \city{Shanghai}
  \country{China}
}
\email{hxliu26@m.fudan.edu.cn}

\author{Lei Zhou}
\affiliation{%
  \institution{Fudan University}
  \city{Shanghai}
  \country{China}
}
\email{zhoulei26@m.fudan.edu.cn}

\author{Yuhao Ren}
\affiliation{%
  \institution{Fudan University}
  \city{Shanghai}
  \country{China}
}
\email{yhren24@m.fudan.edu.cn}

\author{Yumao Wu}
\affiliation{%
  \institution{Fudan University}
  \city{Shanghai}
  \country{China}
}
\email{yumaowu@fudan.edu.cn}

\author{Zhiang Wang}
\affiliation{%
  \institution{Fudan University}
  \city{Shanghai}
  \country{China}
}
\email{zhiangwang@fudan.edu.cn}

\begin{document}

\begin{abstract}
Physical design algorithms operate within tightly coupled, multi-stage optimization flows, where stage-local gains may vanish or induce downstream degradation.
Existing program-evolution frameworks often rely on stage-local
objectives or undifferentiated multi-metric feedback, which neither
guarantee better final results nor identify which unmet requirement
should guide the next iteration.
We present GoalEvolve, a goal-driven framework that makes physical design algorithm evolution accountable for the final quality of results (QoR) of the complete flow.
Given a multi-objective QoR target region, GoalEvolve converts unmet requirements into normalized target gaps, identifies the dominant bottleneck, and uses stage-resolved checkpoint evidence to locate the responsible stage.
An LLM-based Teacher then narrows the search to a relevant algorithmic decision and source region, while parallel Student agents implement and validate hypotheses through full-flow evaluation.
Local effects, optimization debt, and downstream retention are retained as mechanism evidence for subsequent evolution.
Across eight ASAP7 designs, GoalEvolve improves post-route TNS by 30.67\% on average and reduces leakage and dynamic power by 21.18\% and 9.42\% versus default OpenROAD. Relative to commercial-tool goals, it closes 62.20\% of the normalized power gap on power-dominant designs, surpasses the TNS goals on both timing-dominant designs, and closes 32.48\% of the equal-weight timing--power gap on joint designs. Across all three designs evaluated against Codex goal mode under matched budgets, GoalEvolve further improves TNS by 26.46\% while reducing leakage and dynamic power by 12.38\% and 0.76\%, respectively.

\end{abstract}


\maketitle
\section{Introduction}
\label{sec:intro}

Physical design proceeds through a tightly coupled implementation flow
consisting of floorplanning, placement, post-placement optimization,
routing, and post-routing optimization, in which algorithmic decisions
can propagate across stages. A common phenomenon is that an improvement
achieved at one stage does not necessarily translate into better final
QoR, and may even degrade the design after downstream
processing. For example, an aggressive timing transformation after
placement may reduce local violations while increasing congestion,
routing detours, or power, ultimately worsening post-route QoR. A
stage-level decision is therefore valuable only if it preserves design
validity and its benefit persists through the complete flow. This property
imposes a stringent requirement on physical design algorithm evolution:
optimizing an algorithm for a particular stage must account not only for
its immediate objective, but also for its impact on the final
multi-objective QoR of the entire implementation flow.

The conventional development process follows an algorithm-first
paradigm. Researchers typically begin with a perceived weakness,
formulate an optimization strategy, implement the corresponding
mechanism, and then evaluate the complete flow. This approach is
effective when the relationship between the proposed mechanism and the
target QoR is well understood. In many design-closure problems, however,
the desired outcome is explicit while the algorithmic path toward it
remains unclear. Stage logs and intermediate reports provide useful
diagnostic evidence, but translating that evidence into the next
source-level hypothesis still requires substantial manual analysis and
domain expertise. Algorithm development therefore proceeds through
repeated cycles of diagnosis, implementation, and validation, without
an automated way to derive the next intervention directly from the
remaining target gap.

\begin{figure}[t]
  \centering
  \includegraphics[width=0.8\columnwidth]
{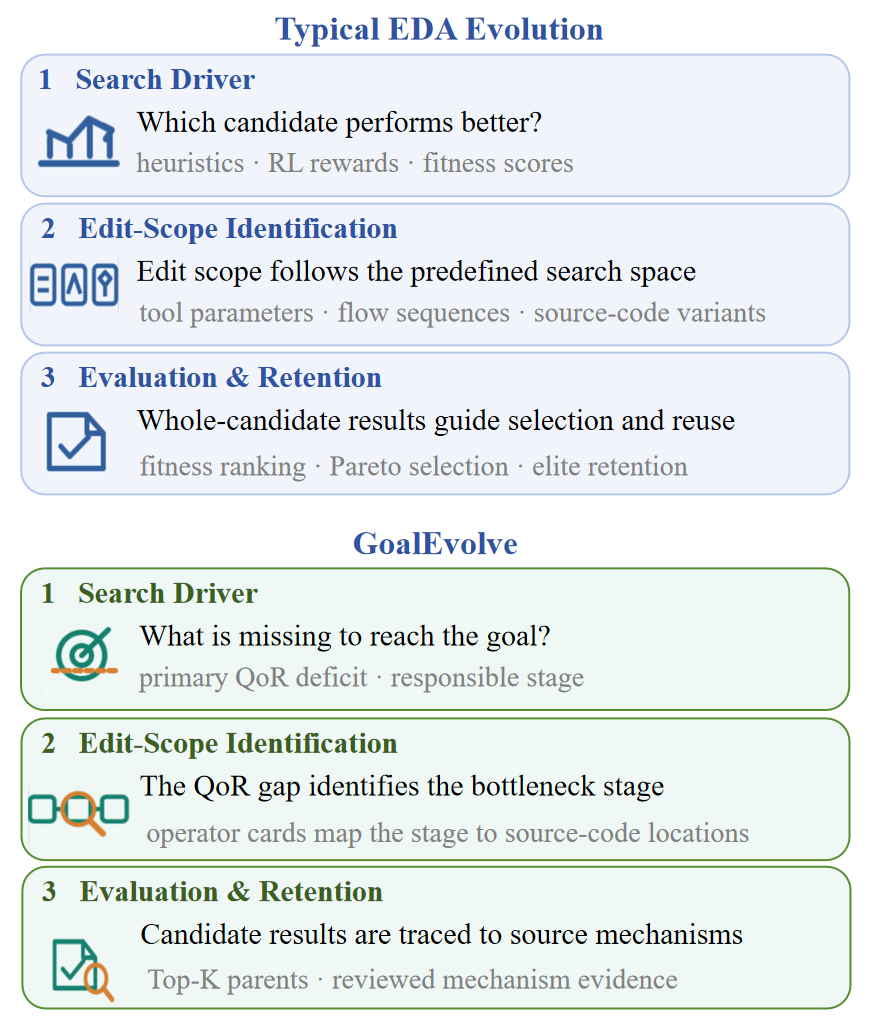}
\caption{Comparison of typical EDA evolution and GoalEvolve across three key dimensions. Unlike score-driven selection, GoalEvolve uses the remaining QoR gap to guide source edits and retain mechanism-level evidence.}
  \label{fig:general_vs_goalevolve}
  \Description{Comparison between general program evolution and GoalEvolve.}
  \vspace{-3pt}
\end{figure}

LLM agents make repository-scale source-code exploration and
modification increasingly practical. General program-evolution systems
combine program generation with automated
evaluation~\cite{romera2024funsearch,liu2024eoh,ye2024reevo,
novikov2025alphaevolve}, while recent EDA systems extend this paradigm
to logic synthesis and multiple stages of physical
design~\cite{yao2025placementevolve,ghose2026audopeda,
jafri2026grevolve,yu2026selfevolvedabc}. These studies show that LLM
agents can iteratively improve EDA algorithms through execution
feedback. Nevertheless, their control logic leaves two issues unresolved
in tightly coupled physical design flows. First, when candidate
selection is based on the objective of the modified stage, an apparent
local gain may be lost or overturned by later stages and thus need not
improve the full-flow outcome. Second, multiple reported metrics are
often presented without an explicit priority. Such feedback can compare
candidates, but does not indicate which remaining target deficit should
govern the next source revision. As a result, the agent may alternate
among competing objectives or modify mechanisms unrelated to the current
bottleneck, leading to an unfocused search and unnecessary full-flow
evaluations.
These limitations suggest that physical design algorithm evolution must
be accountable for final full-flow QoR rather than for improvements
observed only at the stage being modified. Under this full-flow
accountability, QoR should serve not merely as a terminal fitness value,
but as a control signal that determines what the evolution process
should address next. Rather than asking whether a candidate is locally
or globally better than its predecessor, the controller should identify
the final QoR requirement with the largest unresolved target gap, locate
where the gap is introduced or left unrepaired, and focus the next
source modification on the corresponding stage and mechanism. This
changes multi-objective evolution from an undirected search over
aggregate improvements into a target-directed process focused on a
specific unresolved final-QoR requirement in each iteration. Figure~\ref{fig:general_vs_goalevolve}
summarizes this shift from score-driven selection to goal-gap-guided
source evolution.

Based on this principle, we present GoalEvolve, a goal-driven framework
that evolves physical design algorithms toward an explicit
multi-objective QoR target region. Given a QoR specification, GoalEvolve
converts unmet final-QoR requirements into normalized target gaps: the
dominant gap determines \emph{what} to improve, while stage-resolved
checkpoint evidence identifies the responsible stage by showing
\emph{where} the deficiency is introduced or left unrepaired. Guided
by this diagnosis, an LLM-based Teacher narrows the search to the
relevant algorithmic decision and bounded source region, while parallel
Student agents implement and validate hypotheses. Stage contracts relate
each candidate's expected local effect to temporary optimization debt
and specify the downstream evidence needed to verify that the benefit
persists through the complete flow. Successful and failed experiments
are retained as persistent mechanism evidence to refine promising
mechanisms and avoid ineffective interventions. The same controller
operates across designs, while each algorithmic path adapts to
design-specific targets and observed behavior. GoalEvolve thus provides
a reusable methodology for deriving and revising optimization logic
from explicit goals and full-flow evidence rather than improving a
single optimization algorithm.

The main contributions of this work are summarized as follows:
\begin{itemize}[noitemsep, topsep=0pt, leftmargin=*]
\item We introduce goal-driven physical design algorithm evolution toward
explicit multi-objective QoR regions, formalized as target-set
reachability over feasible program variants and evaluated by final
full-flow QoR rather than stage-local gains.

\item We develop a bottleneck-directed loop that identifies the dominant
unmet QoR requirement from normalized target gaps, uses stage-resolved
checkpoints to locate the responsible stage, and narrows the next
intervention to the relevant algorithmic decision and bounded source
region.

\item We establish evidence-constrained hypotheses and persistent
mechanism memory, retaining successful and failed experiments to guide
subsequent evolution.

\item We instantiate GoalEvolve for OpenROAD post-placement optimization.
Across eight ASAP7 designs, it improves post-route TNS by 30.67\% and
reduces leakage and dynamic power by 21.18\% and 9.42\% over default
OpenROAD. Against commercial-tool targets, it closes 62.20\% of the power gap on power-dominant designs and 32.48\% of the joint
timing--power gap, while surpassing TNS targets on both timing-dominant
designs. Under matched budgets on three designs, it further improves
TNS by 26.46\% over Codex goal mode.

\item We will open-source the GoalEvolve implementation and experimental
scripts to support reproducibility and future research.
\end{itemize}

The remainder of this paper is organized as follows. Section~\ref{sec:related} reviews related work on physical design optimization and LLM-based algorithm evolution. Section~\ref{sec:method} presents the GoalEvolve framework for evolving physical design algorithms toward explicit QoR targets and its post-placement instantiation. Section~\ref{sec:experiments} reports the experimental validation, and Section~\ref{sec:conclusion} concludes the paper.

\section{Background and Related Work}
\label{sec:related}

This section traces the progression from expert-driven physical design algorithm development to LLM-based program evolution and repository-scale EDA agents. We first review how representative physical design algorithms rely on domain expertise for development and adaptation, and then examine how evolutionary and agentic systems generate, evaluate, and retain source variants.

\subsection{Expert-Driven Algorithm Engineering}

Recent physical design algorithms have largely advanced through expert-led revision of optimization models and control logic. RePlAce introduces constraint-oriented local-density penalties and dynamic step-size adaptation, whereas DREAMPlace maps analytical placement to PyTorch tensor operators and custom GPU kernels~\cite{cheng2018replace,lin2019dreamplace}. Differentiable timing-driven placement models STA propagation as a differentiable computation to optimize TNS and WNS directly, while FusionSizer jointly optimizes cell locations and gate sizes through continuous relaxation~\cite{guo2022differentiable,du2024fusionsizer}. OpenPhySyn organizes physical synthesis transformations into violation-driven repair procedures, whereas CUGR combines a probabilistic resource model with three-dimensional routing to improve detailed routability~\cite{agiza2020openphysyn,liu2020cugr}. RL-Sizer formulates sequential gate sizing as reinforcement learning driven by timing rewards, while DAGSizer uses directed timing graphs to predict discrete sizing assignments~\cite{lu2021rlsizer,cheng2023dagsizer}. Both methods require task-specific model training within predefined sizing spaces. These advances improve physical design optimization, but their development and adaptation still require substantial backend design expertise.


\subsection{Agentic AI for Algorithm Evolution}

The emergence of LLMs has opened a new path for general algorithm evolution by closing the loop among search-space exploration, code generation, and executable evaluation. FunSearch searches over implementations of a user-specified function within a fixed program scaffold, retains valid programs in island populations, and samples higher-scoring variants for subsequent prompts~\cite{romera2024funsearch}. AlphaEvolve directly edits code and uses an evolutionary database to select parent programs according to feedback from multiple evaluators~\cite{novikov2025alphaevolve}. EoH co-evolves natural-language ideas and executable heuristics through fitness-based population updates, whereas ReEvo converts pairwise performance differences into immediate and accumulated reflections that guide crossover and elitist mutation~\cite{liu2024eoh,ye2024reevo}. LLaMEA retains the best algorithm in a single-parent loop and feeds runtime fitness and errors back for mutation or redesign~\cite{vanstein2024llamea}. ShinkaEvolve improves sample efficiency through exploration-aware parent sampling, novelty filtering, and adaptive LLM selection~\cite{lange2025shinkaevolve}. Together, these systems demonstrate that LLM agents can sustain empirical algorithm evolution through execution feedback and structured retention, providing a practical foundation for extending this paradigm to EDA.

\subsection{Agentic EDA and Source Evolution}

Building on these advances, Agentic EDA extends program evolution to repository-scale tool development and design-flow optimization. Yao et al. use an LLM genetic flow to evolve three predefined components of global placement, selecting candidates by both placement quality and diversity~\cite{yao2025placementevolve}. AuDoPEDA combines repository graphs with literature-grounded planning to localize OpenROAD edits, then uses hard gates and staged flow evaluation to commit non-regressing patches while returning failures as planning counterexamples~\cite{ghose2026audopeda}. GR-Evolve externalizes router variants and QoR histories as versioned artifacts, then selects among bounded-search candidates using Pareto priorities over downstream routing QoR and runtime~\cite{jafri2026grevolve}. Self-Evolved ABC assigns major synthesis subsystems to specialized agents, while a central planner selects the next subsystem from multidimensional QoR feedback, merges improvements into a champion version, and rolls back regressions~\cite{yu2026selfevolvedabc}.  Existing work already uses structured QoR feedback to localize and revise source changes. However, we are not aware of prior work that formalizes the residual to a predefined end-of-flow QoR region as the control signal for assigning stage responsibility and generating the next source-level algorithm hypothesis. GoalEvolve addresses this gap by making the remaining target distance an explicit basis for subsequent evolution.



\section{Our Approach}
\label{sec:method}

\begin{figure*}[t]
  \centering
  \includegraphics[width=0.99\textwidth]{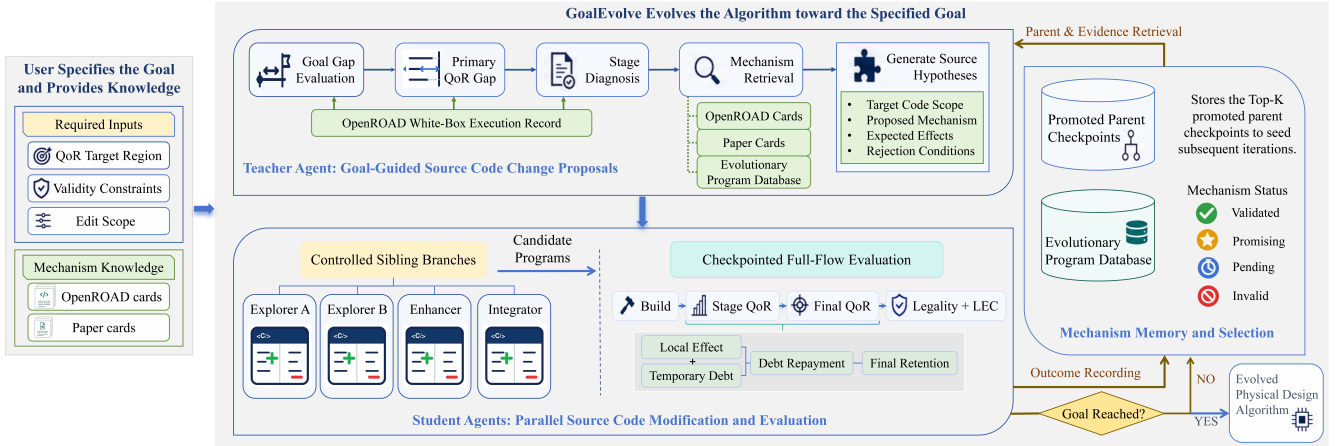}
\caption{Overview of GoalEvolve. The Teacher converts QoR gaps and retrieved knowledge into targeted source hypotheses, while parallel Students evaluate candidates through checkpointed full-flow runs. Reviewed outcomes update mechanism evidence and parent selection for subsequent iterations.}
  \label{fig:overview-v2}
  \Description{GoalEvolve contains goal-guided bottleneck localization, structured evidence retrieval, controlled source experiments, checkpointed full-flow evaluation, a mechanism evidence registry, and a checkpoint portfolio.}
\end{figure*}

To move beyond algorithm-first development, we introduce GoalEvolve,
a framework for goal-driven evolution of physical design algorithms.
It couples two feedback loops: an optimization loop that translates
the remaining target gap into a bounded source code experiment, and
an evidence loop that converts completed experiments into persistent
mechanism knowledge. The remaining goal gap therefore acts as an active
control signal for deciding what evidence to inspect and which source
mechanism to revise, rather than serving only as a terminal score.
Figure~\ref{fig:overview-v2} provides a structural overview, while
Algorithm~\ref{alg:goalevolve} summarizes the corresponding execution
loop.

\subsection{Goal-Guided Evolution Planning}
\label{subsec:diagnosis}

GoalEvolve evaluates a program variant $p$ on a design $d$ against
design-specific QoR targets. After orienting all metrics to a lower-is-better form, with
$|\mathrm{TNS}|$ used for TNS, we define the normalization scale,
normalized target violation, and overall goal distance as follows:

\begin{gather}
s_i(d)=
\max\!\left\{
\left|q_i(p_0,d)-u_i(d)\right|,
\epsilon_i
\right\},
\label{eq:normalization-scale}
\\
r_i(p,d)=
\left[
\frac{q_i(p,d)-u_i(d)}{s_i(d)}
\right]_{+},
\label{eq:target-violation}
\\
D_F(p,d)=
\sum_{i=1}^{m} w_i r_i(p,d),
\label{eq:goal-distance-v2}
\end{gather}

\noindent where $[x]_{+}\equiv\max(0,x)$, $m$ denotes the
number of selected QoR metrics, and $w_i>0$ is the weight assigned
to metric $i$. Here, $p_0$ denotes the baseline program, $q_i(p,d)$ is
the QoR value measured by the fixed evaluation flow, and $u_i(d)$ is
the corresponding inclusive upper target bound for design $d$. The
positive $\epsilon_i$ prevents division by zero
when the baseline value equals the target. Consequently, $D_F(p,d)=0$ if and only if all
selected QoR targets are satisfied. A candidate is deemed goal-compliant
only if it also passes all required validity checks.

\begin{algorithm}[t]
\caption{Core execution loop of GoalEvolve.}
\label{alg:goalevolve}
\footnotesize
\begin{algorithmic}[1]
\Require Initial program $p_0$, design $d$, frozen goal $G$, fixed
evaluation flow $F$, OpenROAD Cards $\mathit{openroadCards}$,
Paper Cards $\mathit{paperCards}$, and evaluation budget $B$

\Ensure Best feasible OpenROAD program $p_{\mathrm{best}}$

\Variables $\mathrm{EPD}[p]$ stores the evaluation record and mechanism
status of program $p$; \textit{pool} retains feasible programs across
rounds; \textit{parent} is inherited by the next round, whereas
$p_{\mathrm{best}}$ is the current champion; \textit{diagnosis},
\textit{mechanisms}, and \textit{hypotheses} denote the identified
bottleneck, retrieved algorithmic options, and planned branch
modifications, respectively; $h$ denotes one hypothesis and
\textit{child} its implementation.

\State $\mathrm{EPD}\gets\emptyset$
\State $\mathrm{EPD}[p_0]\gets\Call{Evaluate}{p_0,d,F}$
\State $\mathit{pool}\gets\{p_0\}$;
$\mathit{parent}\gets p_0$;
$p_{\mathrm{best}}\gets p_0$

\While{$\Call{Remaining}{B}$ \textbf{ and not }
$\Call{Satisfy}{\mathrm{EPD}[p_{\mathrm{best}}],G}$}
    \State $\mathit{diagnosis}\gets
    \Call{Diagnose}
    {\mathrm{EPD}[\mathit{parent}],G,\mathit{openroadCards}}$
    \State $\mathit{mechanisms}\gets
    \Call{Retrieve}
    {\mathit{diagnosis},\mathit{paperCards},\mathrm{EPD}}$
    \State $\mathit{hypotheses}\gets
    \Call{Plan}{\mathit{diagnosis},\mathit{mechanisms}}$

    \ForAll{$h\in\mathit{hypotheses}$ \textbf{ in parallel}}
        \State $\mathit{child}\gets
        \Call{Implement}{\mathit{parent},h}$
        \State $\mathrm{EPD}[\mathit{child}]\gets
        \Call{Evaluate}
        {\mathit{child},d,F,\mathrm{EPD}[\mathit{parent}]}$
        \State $\mathrm{EPD}\gets
        \Call{Update}{\mathrm{EPD},h,\mathit{child}}$
        \State $\mathit{pool}\gets
        \Call{Retain}
        {\mathit{pool},\mathit{child},\mathrm{EPD}[\mathit{child}]}$
    \EndFor

    \State $p_{\mathrm{best}}\gets
    \Call{Champion}{\mathit{pool},\mathrm{EPD}}$
    \State $\mathit{parent}\gets
    \Call{Select}
    {p_{\mathrm{best}},\mathit{pool},\mathit{diagnosis},\mathrm{EPD}}$
    \State $\mathit{openroadCards}\gets
    \Call{Refresh}
    {\mathit{openroadCards},\mathit{parent},
    \mathrm{EPD}[\mathit{parent}]}$
\EndWhile

\State \Return $p_{\mathrm{best}}$
\end{algorithmic}
\end{algorithm}

After each round, GoalEvolve identifies the dominant QoR bottleneck from
the weighted target gaps and traces it through the white-box checkpoint
trajectory. By mapping final metrics to their affecting stages, it
compares expected and observed stage effects while accounting for
unresolved optimization debt and downstream loss of earlier gains. A
compact reflection prompt links the dominant gap to relevant checkpoint
and object-level evidence, retrieving supporting artifacts by reference
only when needed. This converts the global gap into a focused evidence
request for the next source modification.

The Teacher retrieves potential mechanisms from three complementary
sources. OpenROAD Cards map stage behavior to implementation locations
and preservation constraints. Paper Cards translate published methods
into implementable mechanisms with expected effects and rejection
criteria. Historical records from the Evolutionary Program Database
in Section~\ref{subsec:memory} help avoid ineffective mechanisms and
reconsider promising ones.

Guided by the diagnosis and retrieved evidence, the Teacher prioritizes
mechanisms most likely to address the responsible stage, balancing new
exploration with refinement or integration supported by prior evidence.
Each hypothesis links a concrete source modification to its expected
stage response and final QoR effect, with a rejection criterion for
retention. GoalEvolve thereby narrows an unmet QoR target to a
responsible stage, candidate mechanism, and source implementation,
which together define the next evolution target.
Lines~5--7 of Algorithm~\ref{alg:goalevolve} summarize this
diagnosis--retrieval--planning sequence.

\subsection{Controlled Evolution and Attribution}
\label{subsec:controlled}

GoalEvolve evaluates source hypotheses through four controlled Student
branches that start from the same parent version. Two Explorer branches
implement distinct new ideas, one Enhancer strengthens a mechanism whose
gains are promising but not yet stable, and one Integrator combines
previously validated mechanisms when their source modifications and
expected stage effects are compatible. The common parent and evaluation
flow provide a consistent reference for attributing each branch outcome
to its assigned hypothesis.

Within each branch, an LLM coding agent with direct repository access and command execution capabilities implements the assigned hypothesis and builds an isolated tool instance. When compilation fails or another implementation issue arises, the agent autonomously diagnoses the problem and revises the patch within a fixed repair budget, preventing an engineering error from prematurely rejecting the underlying mechanism.

GoalEvolve introduces effect--debt analysis, which defines the immediate change in the target QoR metric as the local effect and degradation introduced in other metrics as optimization debt. Both quantities are measured against the common parent after the target stage, while the final comparison measures effect retention and debt repayment. GoalEvolve also tracks object-level netlist changes across checkpoints to identify operations that are later repeated or reversed and therefore require further evolution. These metric and object traces form an evidence signature that binds the source diff to observed behavior for subsequent mechanism selection.
Lines~8--13 of Algorithm~\ref{alg:goalevolve} summarize this
controlled evolution-and-attribution process.

\subsection{Mechanism Memory and Selection}
\label{subsec:memory}

GoalEvolve builds an Evolutionary Program Database (EPD), which stores all ideas proposed by the Teacher together with an idea tag identifying the optimization target, evaluation records, source patches, and other accumulated feedback. Each record is labeled as \emph{validated}, \emph{promising}, \emph{pending}, or \emph{invalid} according to its evaluation status and outcome, enabling the EPD to balance exploitation and exploration. For exploitation, compatible validated mechanisms are assigned to the Integrator, while the Enhancer can perform at most $K_{\mathrm{rep}}$
bounded repairs on a promising record. The EPD also retains programs based on distinct mechanisms even when they do not achieve the highest current scores, preventing evolution from being confined to a single path. In parallel, Explorers implement and evaluate newly proposed
algorithmic ideas with no matching record in the database. Negative evaluation evidence suppresses repeated proposals and updates the contraindications in mechanism cards. As the database grows, compatibility with the responsible stage and current source state first narrows the candidate set, after which the LLM agent reasons over feedback from previous rounds to select the most promising mechanisms for enhancement or integration.

The evidence accumulated in the EPD also guides checkpoint inheritance. During regular evolution, sibling experiments inherit the champion $p^*$, defined as the feasible checkpoint with the smallest goal distance, while GoalEvolve retains a top-$K$ portfolio containing $p^*$ and feasible alternatives from distinct mechanism families. A plateau is detected when the best goal distance fails to improve beyond a preset threshold for several consecutive rounds. GoalEvolve then samples the next parent $c$ from this portfolio according to

\begin{equation}
 P_j(c)\propto
 \exp\!\left(-\frac{D_F(c)-D_F(p^*)}{T_j}\right)
 (1+\beta N_c)(1+\gamma A_c),
 \label{eq:parent-v2}
\end{equation}

where $N_c$ measures mechanism diversity and $A_c$ measures alignment with the current bottleneck:

\begin{equation}
 N_c=1-\frac{h(\phi_c)}{H},
 \qquad
 A_c=M_{\tau_b,\tau_c}.
 \label{eq:parent-factors}
\end{equation}

Here, $h(\phi_c)$ counts how often the mechanism family $\phi_c$ was selected in the previous $H$ inheritance decisions. The entry $M_{\tau_b,\tau_c}\in[0,1]$ of a fixed tag compatibility matrix gives the compatibility between the candidate tag $\tau_c$ and the bottleneck tag $\tau_b$, while $\beta$ and $\gamma$ weight the diversity and alignment factors. At inheritance iteration $j$ after plateau detection, GoalEvolve normalizes $P_j(c)$ over the top-$K$ portfolio and samples one checkpoint. The temperature $T_j$ is gradually reduced according to a fixed schedule as $j$ increases. The champion is retained throughout this process, allowing GoalEvolve to explore another inheritance path without losing the best program found so far.

\subsection{Post-Placement Instantiation}
\label{subsec:postplace}

We instantiate GoalEvolve for post-placement timing and power
optimization, where gate resizing, buffering, and VT reassignment
improve a placed netlist, but routing-extracted RC parasitics may erode
these gains. This gap between immediate improvement and downstream
retention makes post-placement optimization a representative full-flow
case study. All candidates start from the same placed netlist, while
timing constraints, technology data, downstream routing, and the final
evaluator remain fixed. The editable source scope is limited to Gate
Resizer and Restructure mechanisms controlling candidate generation,
scoring, scheduling, and commit decisions. The goal contract specifies post-route timing and power targets,
while runtime is recorded as an observer metric.
White-box telemetry and structural statistics, including buffer growth
and VT composition, provide diagnostic evidence.

Each full-flow evaluation records checkpoints after power reclaim,
timing recovery, and supplemental repair. These observation contracts
expose local effects, optimization debt, and final retention, allowing
the Teacher to prioritize the next stage. The initial curriculum applies power reclaim before timing recovery,
while power-oriented or timing-oriented restructuring mechanisms
become eligible only after cell-level optimization reaches a measured
plateau. This instantiation maps GoalEvolve to a concrete source
evolution task without imposing this stage sequence on the general
framework.

\begin{table*}[t]
\centering
\footnotesize
\setlength{\tabcolsep}{2pt}
\renewcommand{\arraystretch}{1.15}
\caption{Frozen-contract baseline, goal, and recorded GoalEvolve result
(Baseline/Goal/Achieved). Runtime is an observer metric with a recorded baseline but no frozen goal.}
\label{tab:goal-attainment}

\begin{tabular*}{\textwidth}{
@{\extracolsep{\fill}}c|
r@{\hspace{0.15em}/\hspace{0.15em}}r@{\hspace{0.15em}/\hspace{0.15em}}r|
r@{\hspace{0.15em}/\hspace{0.15em}}r@{\hspace{0.15em}/\hspace{0.15em}}r|
r@{\hspace{0.15em}/\hspace{0.15em}}r@{\hspace{0.15em}/\hspace{0.15em}}r|
r@{\hspace{0.15em}/\hspace{0.15em}}r@{\hspace{0.15em}/\hspace{0.15em}}r@{}
}
\hline
Design &
\multicolumn{3}{c|}{$\mathrm{TNS}$ (ns)} &
\multicolumn{3}{c|}{$P_{\mathrm{leak}}$ ($\mu\mathrm{W}$)} &
\multicolumn{3}{c|}{$P_{\mathrm{dyn}}$ ($\mu\mathrm{W}$)} &
\multicolumn{3}{c@{}}{Runtime (s)} \\
\hline

AES &
$-12.44$ & $\mathbf{-12.00}$ & $-15.57$ &
96.3 & 35.0 & \textbf{29.1} &
431{,}903.7 & 350{,}000.0 & \textbf{335{,}607.1} &
32 & -- & 235 \\

Ariane &
$-7{,}568.43$ & $-1{,}850.00$ & $\mathbf{-688.54}$ &
17{,}900.0 & \textbf{17{,}600.0} & 17{,}935.1 &
640{,}100.0 & 623{,}000.0 & \textbf{588{,}058.6} &
687 & -- & 378 \\

JPEG &
$\mathbf{-48.77}$ & $-53.00$ & $-51.18$ &
174.0 & \textbf{80.0} & 114.5 &
293{,}826.0 & \textbf{250{,}000.0} & 270{,}427.7 &
187 & -- & 745 \\

MemPool &
$-3{,}680.33$ & $\mathbf{-1{,}850.00}$ & $-2{,}714.69$ &
3{,}070.0 & \textbf{1{,}210.0} & 3{,}059.5 &
275{,}930.0 & \textbf{185{,}000.0} & 249{,}586.4 &
1{,}415 & -- & 489 \\

NVDLA-A &
$-203.23$ & $-294.00$ & $\mathbf{-90.64}$ &
321.0 & \textbf{46.3} & 285.3 &
164{,}679.0 & \textbf{126{,}000.0} & 144{,}228.6 &
166 & -- & 828 \\

NVDLA-C &
$-55.58$ & $-10.00$ & $\mathbf{-7.25}$ &
17{,}300.0 & \textbf{16{,}500.0} & 17{,}259.4 &
583{,}700.0 & \textbf{583{,}000.0} & 584{,}086.0 &
632 & -- & 124 \\

NVDLA-M &
$-13.99$ & $\mathbf{-10.40}$ & $-14.21$ &
61.5 & \textbf{31.3} & 38.3 &
44{,}538.5 & \textbf{35{,}300.0} & 39{,}473.3 &
46 & -- & 526 \\

NVDLA-P &
$-198.18$ & $-246.00$ & $\mathbf{-163.49}$ &
306.0 & \textbf{66.6} & 256.4 &
39{,}694.0 & \textbf{36{,}400.0} & 38{,}231.5 &
125 & -- & 1{,}074 \\

\hline
\end{tabular*}
\end{table*}

\section{Experiments}
\label{sec:experiments}

This section evaluates GoalEvolve's post-route QoR improvement and
progress toward frozen commercial-tool targets across eight ASAP7
designs. We then validate whether these gains arise from evolved
source-level mechanisms and examine their transfer across designs.
Finally, we compare GoalEvolve with Codex goal mode under matched
formal-evaluation budgets and analyze the evolved JPEG mechanisms in
detail.

\subsection{Experimental Setup}
\label{subsec:setup}

All experiments are conducted on a Rocky Linux 8.10 server with two
Intel Xeon Platinum 8462Y+ processors and 314\,GiB of memory. We evaluate
GoalEvolve on eight post-placement designs spanning approximately
20\,K--313\,K instances. All designs use the ASAP7 7\,nm predictive process design kit and its
7.5-track standard-cell library~\cite{clark2016asap7,vashishtha2017asap7}. Our implementation is based on
OpenROAD~\cite{ajayi2019openroad} commit \texttt{08f67ee5}.

Target construction is performed before evolution and is external to
GoalEvolve. For each design, \textbf{we use preliminary experiments with
a commercial reference flow to characterize attainable post-placement
repair opportunities and set its post-route TNS, leakage, and dynamic-power
results as the frozen evolution targets.} These targets are fixed before
the first GoalEvolve or Codex goal mode run and reported in
Table~\ref{tab:goal-attainment}; runtime is recorded only as an observer
metric.

To ensure controlled evaluation, all candidate branches start from the
same parent source snapshot and design checkpoint, but are built in
isolated workspaces and evaluated with an identical checkpointed
post-placement and global-routing flow. A GoalEvolve round comprises one
Teacher and four coding agents, all using \texttt{GPT-5.6 Terra} with high reasoning effort. Evolution terminates when the predefined goal is
reached or the full-flow evaluation budget is exhausted.


\subsection{QoR Improvement and Goal Progress}
\label{subsec:goal-attainment}

Table~\ref{tab:goal-attainment} reports the OpenROAD baseline $B$,
commercial-tool goal $G$, and final GoalEvolve result $A$.
To characterize the repair opportunities implied by the commercial-tool
targets, we derive baseline-to-goal timing and total-power improvement
potentials. Table~\ref{tab:goal-profiles} groups the eight designs into
four power-dominant, two timing-dominant, and two joint timing--power
cases based on these potentials. This grouping is used only for the
result analysis below. We measure the fractions of the baseline-to-goal
timing and total-power gaps closed by GoalEvolve using $C_T$ and $C_P$,
respectively, and use their equal-weight combination
$C_J=(C_T+C_P)/2$ for joint timing--power cases:
\begin{equation}
C_T=
\min\!\left(
1,\,
\frac{|\mathrm{TNS}_B|-|\mathrm{TNS}_A|}
     {|\mathrm{TNS}_B|-|\mathrm{TNS}_G|}
\right).
\label{eq:timing-gap-closure}
\end{equation}
\begin{equation}
C_P=
\min\!\left(
1,\,
\frac{P_{\mathrm{tot},B}-P_{\mathrm{tot},A}}
     {P_{\mathrm{tot},B}-P_{\mathrm{tot},G}}
\right).
\label{eq:power-gap-closure}
\end{equation}

\begin{table}[t]
\centering
\footnotesize
\setlength{\tabcolsep}{3pt}
\renewcommand{\arraystretch}{1.08}
\caption{Baseline-to-goal improvement potentials and corresponding QoR profiles.}
\label{tab:goal-profiles}
\begin{tabular*}{\columnwidth}{
@{\extracolsep{\fill}}cccc@{}
}
\toprule
Design &
Timing potential &
Total-power potential &
Profile \\
\midrule
AES     & +3.54\%  & +18.97\% & Power-dominant \\
Ariane  & +75.56\% & +2.64\%  & Timing-dominant \\
JPEG    & -8.67\%  & +14.94\% & Power-dominant \\
MemPool & +49.73\% & +33.26\% & Joint \\
NVDLA-A & -44.66\% & +23.61\% & Power-dominant \\
NVDLA-C & +82.01\% & +0.25\%  & Timing-dominant \\
NVDLA-M & +25.66\% & +20.78\% & Joint \\
NVDLA-P & -24.13\% & +8.83\%  & Power-dominant \\
\bottomrule
\end{tabular*}
\end{table}

\begin{figure}[t]
  \centering
  \includegraphics[width=1\columnwidth]{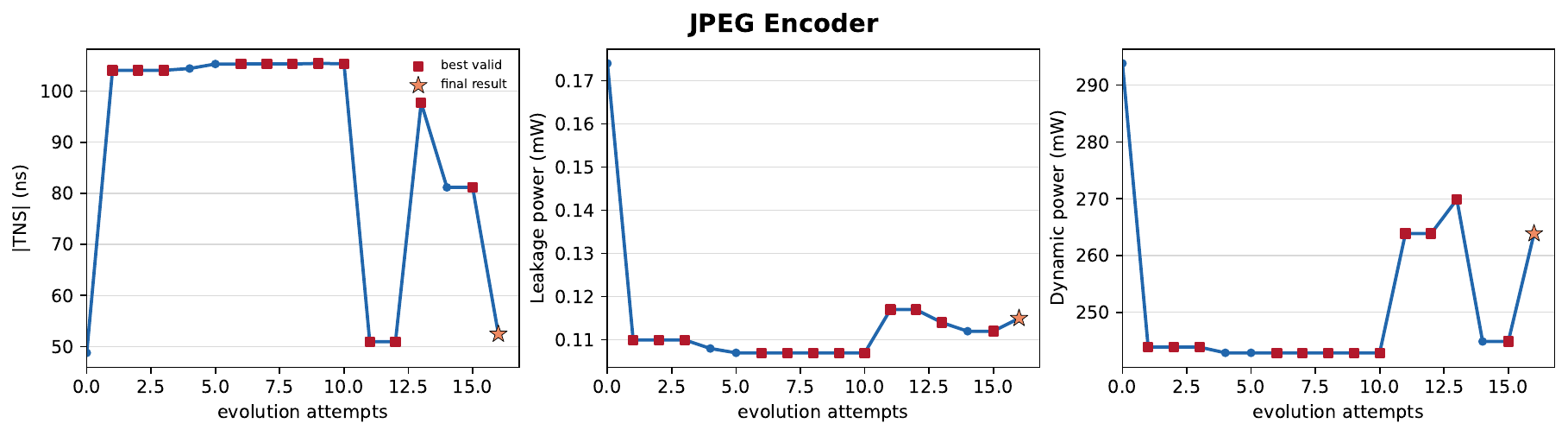}
  \vspace{1mm}
  \includegraphics[width=1\columnwidth]{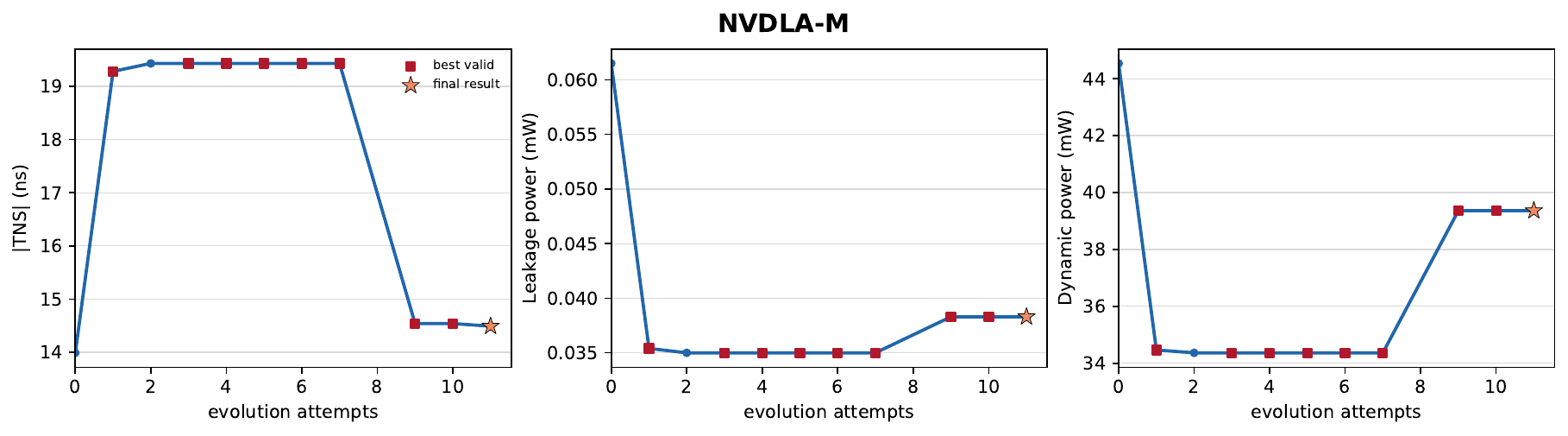}
\caption{Timing and power trajectories across evolution attempts for
JPEG (top) and NVDLA-M (bottom).}
  \label{fig:evolution-trajectories}
\end{figure}

Relative to the OpenROAD baseline, GoalEvolve improves post-route TNS by
30.67\% on average while reducing leakage and dynamic power by 21.18\%
and 9.42\%, respectively. The timing improvement is substantial on
several large-gap designs: Ariane, NVDLA-C, and NVDLA-A each improve TNS
by at least 55.40\%. The three designs with TNS regressions incur at most
3.13\,ns of degradation while still reducing leakage by 34.20--69.78\%
and dynamic power by 7.96--22.30\%. Relative to the commercial-tool goals,
GoalEvolve closes 62.20\% of the normalized total-power gap on the four
power-dominant designs and surpasses the commercial-tool TNS on three of
them. On the two timing-dominant designs, GoalEvolve surpasses the
commercial-tool TNS on both while achieving 2.55\% lower total power on
average. For the two joint timing--power designs, it closes 32.48\% of
the equal-weight normalized timing--power gap. 
Figure~\ref{fig:evolution-trajectories} further shows the timing and power trajectories
of JPEG and NVDLA-M, revealing the non-monotonic timing--power tradeoffs
across successive evolution attempts. Together with the endpoint results,
these trajectories show that GoalEvolve can navigate different QoR
tradeoffs and make substantial progress toward the commercial-tool targets.

\subsection{Validation of Evolved Source Mechanisms}
\label{subsec:source-validation}

The repeated-baseline control is applied to the five designs for which
GoalEvolve has substantially higher runtime than the default flow,
allowing multiple baseline passes within the corresponding runtime
budget. To test whether the gains arise merely from additional
optimization time, we give the repeated baseline the largest integer
number of default passes that fits within GoalEvolve's recorded runtime,
corresponding to 0.75--0.96$\times$ the GoalEvolve budget.
Table~\ref{tab:repeated-baseline} shows that GoalEvolve improves TNS
by 3.09\% on average while reducing leakage and dynamic power by
40.94\% and 14.59\%, respectively. Repeated execution of the existing
optimization operators within a comparable runtime budget therefore
does not reproduce the QoR gains of the evolved source.

To further examine whether these evolved mechanisms are specific to
the design on which they were discovered, we freeze the final
AES-evolved OpenROAD source and replay it on the other seven designs
using the baseline flow, without further source evolution. For this
comparison, we compute $D_F$ using Equation~\ref{eq:goal-distance-v2}
with equal weights on TNS, leakage, and dynamic power.
Table~\ref{tab:cross-design-transfer} shows that the transferred source
improves TNS on all seven designs by 46.72\% on average and reduces
total power on six of them. Nevertheless, design-specific GoalEvolve achieves
a smaller $D_F$ on six of the seven designs. NVDLA-C is the only
exception, where the transferred source achieves $D_F=0.8176$ versus
$0.8336$ for design-specific evolution. These results indicate that the
evolved source mechanisms exhibit useful cross-design transfer, while
design-specific goal feedback remains important for approaching each
frozen QoR target.

\begin{table}[t]
  \centering
  \caption{GoalEvolve (GE) vs. repeated baseline (RB).}
  \label{tab:repeated-baseline}
  \resizebox{\columnwidth}{!}{
  \begin{tabular}{ccccc}
    \toprule
    Design &
    RB passes &
    $\mathrm{TNS}$ (ns) GE/RB &
    $P_{\mathrm{leak}}$ ($\mu$W) GE/RB &
    $P_{\mathrm{dyn}}$ (mW) GE/RB \\
    \midrule
    AES
      & 7
      & $-15.57/\mathbf{-12.90}$
      & $\mathbf{29.1}/101.0$
      & $\mathbf{335.6}/435.9$ \\
    JPEG
      & 3
      & $-51.18/\mathbf{-41.91}$
      & $\mathbf{114.5}/199.0$
      & $\mathbf{270.4}/299.8$ \\
    NVDLA-A
      & 4
      & $\mathbf{-90.64}/-199.06$
      & $\mathbf{285.3}/337.0$
      & $\mathbf{144.2}/164.7$ \\
    NVDLA-M
      & 11
      & $\mathbf{-14.21}/-18.20$
      & $\mathbf{38.3}/82.2$
      & $\mathbf{39.5}/51.6$ \\
    NVDLA-P
      & 8
      & $-163.49/\mathbf{-138.39}$
      & $\mathbf{256.4}/330.0$
      & $\mathbf{38.2}/39.9$ \\
    \bottomrule
  \end{tabular}}
\end{table}

\begin{table}[t]
  \centering
  \footnotesize
  \setlength{\tabcolsep}{3pt}
  \renewcommand{\arraystretch}{1.08}
  \caption{Cross-design replay of the AES-evolved source.
  Positive $\Delta$ values indicate improvement; lower $D_F$ is better.}
  \label{tab:cross-design-transfer}
  \begin{tabular*}{\columnwidth}{
    @{\extracolsep{\fill}}cccc@{}
  }
    \toprule
    Design &
    $\Delta\mathrm{TNS}$ (\%) &
    $\Delta P_{\mathrm{tot}}$ (\%) &
    $D_{F,\mathrm{AES}}/D_{F,\mathrm{GE}}$ \\
    \midrule
    Ariane
      & +91.10 & +7.86
      & 0.373191/\textbf{0.372333} \\
    JPEG
      & +10.01 & +3.26
      & 0.525724/\textbf{0.277710} \\
    MemPool
      & +22.60 & +11.06
      & 0.732363/\textbf{0.725688} \\
    NVDLA-A
      & +64.75 & +0.80
      & 0.637591/\textbf{0.447106} \\
    NVDLA-C
      & +86.16 & -0.05
      & \textbf{0.817592}/0.833560 \\
    NVDLA-M
      & +8.23 & +6.24
      & 0.751243/\textbf{0.581600} \\
    NVDLA-P
      & +44.19 & +1.84
      & 0.579678/\textbf{0.449609} \\
    \bottomrule
  \end{tabular*}
\end{table}

\subsection{Comparison with Codex Goal Mode}
\label{subsec:codex-comparison}

To isolate the benefit of structured physical-design feedback, we
compare GoalEvolve with Codex operating in goal mode on Ariane,
NVDLA-M, and NVDLA-P. Both methods start from the same source
snapshot, use \texttt{GPT-5.6 Terra}, and share the same evaluator and
frozen QoR goals. Codex receives these goals in a natural-language
prompt and iterates directly between source revision and full-flow
evaluation, without GoalEvolve's stage-aware diagnosis, mechanism
memory, or evidence-guided planning. We match the number of valid
formal evaluations exactly between the two methods, counting an
evaluation only when the candidate completes the full flow and returns
all required QoR metrics.

Under these matched evaluation budgets, GoalEvolve achieves better QoR
in seven of the nine timing--power comparisons in
Table~\ref{tab:codex-comparison}. Averaged across the three designs,
GoalEvolve improves TNS by 26.46\% relative to Codex goal mode while
reducing leakage and dynamic power by 12.38\% and 0.76\%, respectively.
GoalEvolve improves all three QoR metrics on Ariane and NVDLA-P.
On NVDLA-M, Codex achieves 0.99\,ns better TNS and only 0.25\% lower
dynamic power, while GoalEvolve reduces leakage by 36.17\%. These results
indicate that GoalEvolve's structured physical-design feedback produces
higher-quality source revisions under the same budget of expensive
full-flow evaluations.


\begin{table}[t]
\centering
\footnotesize
\setlength{\tabcolsep}{1.5pt}
\renewcommand{\arraystretch}{1.12}
\caption{GoalEvolve (GE) versus Codex goal mode (CGM).}
\label{tab:codex-comparison}
\begin{tabular*}{\columnwidth}{@{\extracolsep{\fill}}ccccc@{}}
\toprule
\multirow[c]{2}{*}{Design} &
\multicolumn{1}{c}{Valid rounds/evals.} &
\multicolumn{1}{c}{$\mathrm{TNS}$ (ns)} &
\multicolumn{1}{c}{$P_{\mathrm{dyn}}$ (mW)} &
\multicolumn{1}{c}{$P_{\mathrm{leak}}$ ($\mu$W)} \\
& \multicolumn{1}{c}{(GE; CGM)} &
\multicolumn{1}{c}{GE / CGM} &
\multicolumn{1}{c}{GE / CGM} &
\multicolumn{1}{c}{GE / CGM} \\
\midrule
\multirow[c]{2}{*}{Ariane} &
GE: $3$ r, $12$ evals. &
$\mathbf{-688.54}$ & $\mathbf{588.1}$ & $\mathbf{17{,}935.1}$ \\
& CGM: $12$ evals. &
$-2{,}548.99$ & $594.0$ & $18{,}000.0$ \\
\midrule
\multirow[c]{2}{*}{NVDLA-M} &
GE: $10$ r, $40$ evals. &
$-14.21$ & $39.5$ & $\mathbf{38.3}$ \\
& CGM: $40$ evals. &
$\mathbf{-13.22}$ & $\mathbf{39.4}$ & $60.0$ \\
\midrule
\multirow[c]{2}{*}{NVDLA-P} &
GE: $3$ r, $12$ evals. &
$\mathbf{-163.49}$ & $\mathbf{38.2}$ & $\mathbf{256.4}$ \\
& CGM: $12$ evals. &
$-189.81$ & $38.8$ & $258.0$ \\
\bottomrule
\end{tabular*}
\end{table}




\subsection{Case Study: JPEG on ASAP7}
\label{subsec:case}

\textcolor{black}{
On JPEG (ASAP7), GoalEvolve evolves a code-level, stage-aware
power-recovery policy rather than merely retuning a Tcl schedule.
The evolved flow coordinates three functional stages. First, an early
forced power-reclaim stage aggressively explores power-saving cell
changes before major timing recovery, with ties in utility stably
broken in favor of candidates with larger slack. Second, a bounded
second-pass leakage-recovery stage immediately follows the initial
reclaim to capture residual leakage-saving opportunities while
enforcing explicit timing guards. Third, a timing--power tradeoff
stage performs major timing recovery, then reuses restored slack for
limited guarded power reduction before a final timing repair.}

The second-pass leakage-recovery stage also evolves the internal search
and acceptance policy. Its main search uses a proportional move budget
capped at 4,200 moves. Once this budget is exhausted, one or two bounded
tail searches may examine at most 512 additional candidates each and
retain at most 256 guard-passing moves. For guarded power-VT swaps, the implementation recomputes total
cell leakage after parasitic and timing updates and rolls back the
move if the measured leakage does not decrease or its timing budget
is violated. The evolved source therefore combines slack-aware candidate ordering,
staged timing--power coordination, bounded follow-up search,
post-update benefit verification, and guarded rollback. At the final post-route
endpoint, leakage decreases from 0.1740 to 0.1145\,mW (34.2\%) and
dynamic power from 293.826 to 270.428\,mW (8.0\%), while TNS reaches
$-51.175$\,ns, surpassing the frozen $-53$\,ns target.

\section{Conclusion}
\label{sec:conclusion}

GoalEvolve reframes physical design algorithm development as
goal-directed source evolution under full-flow feedback. Remaining QoR
gaps identify the stage and source mechanism to revise, while checkpoint
evidence and mechanism memory favor effects that survive downstream
interactions. Experiments show that GoalEvolve improves post-route QoR
over default OpenROAD, makes substantial progress toward commercial-tool
targets across heterogeneous timing--power profiles, demonstrates useful
cross-design transfer of evolved mechanisms, and achieves better post-route
QoR than Codex goal mode under matched full-flow evaluation budgets. These results establish goal-guided source
evolution as a practical methodology for physical design.


\clearpage
\bibliographystyle{ACM-Reference-Format}
\bibliography{references}

\end{document}